\documentclass[lettersize,journal]{IEEEtran}
\usepackage{amsmath,amsfonts}
\usepackage{indentfirst}
\usepackage{algorithmic}
\usepackage{algorithm}
\usepackage{array}
\usepackage[caption=false,font=normalsize,labelfont=sf,textfont=sf]{subfig}
\usepackage{textcomp}
\usepackage{stfloats}
\usepackage{url}
\usepackage{verbatim}
\usepackage{graphicx}
\usepackage{epstopdf}
\usepackage{setspace}
\usepackage{cite}
\usepackage{amssymb}
\begin{document}

\title{A Covariance Adaptive Student's t Based\\Kalman Filter}

\author{Benyang Gong, Jiacheng He, Gang Wang, Bei Peng
\thanks{This work was founded by the National Natural Science Foundation of China with Grant 51975107 and Sichuan Science and Technology Major Project No. 2022ZDZX0039, No.2019ZDZX0020, and Sichuan Science and Technology Program No. 2022YFG0343.}
\thanks{B. Gong, J. He and B. Peng are with the School of Mechanical and Electrical Engineering, University of Electronic Science and Technology of China, Chengdu, 611731, PR China (e-mail: 2545608229@qq.com; hejiacheng 123@163.com; beipeng@uestc.edu.cn). G. Wang is with the School of Information and Communication Engineering, University of Electronic Science and Technology of China, Chengdu, 611731, PR China (e-mail: wanggang hld@uestc.edu.cn).}}

 %The paper headers
%\markboth{Journal of \LaTeX\ Class Files,~Vol.~14, No.~8, August~2023}%


\maketitle

\begin{abstract}
In the classical Kalman filter(KF), the estimated state is a linear combination of the one-step predicted state and measurement state, their confidence level change when the prediction mean square error matrix and covariance matrix of measurement noise vary. The existing student's t based Kalman filter(TKF) works similarly to the way KF works, they both work well with impulse noise, but when it comes to Gaussian noise, TKF encounters an adjustment limit of the confidence level, this can lead to inaccuracies in such situations. This brief optimizes TKF by using the Gaussian mixture model(GMM), which generates a reasonable covariance matrix from the measurement noise to replace the one used in the existing algorithm and breaks the adjustment limit of the confidence level. At the end of the brief, the performance of the covariance adaptive student's t based Kalman filter(TGKF) is verified.
\end{abstract}


\section{Introduction}
\IEEEPARstart{K}{alman} filter is now widely used in many fields including state estimation, signal processing, navigation, etc. due to the ease of implementation and scalability of its algorithm. Kalman filter is excellent in the environment of linear state space model and Gaussian noise, it can obtain the optimal minimum mean square error\cite{ref1,ref2,ref3}. However, in practical applications as well as in experiments, due to the presence of non-Gaussian noise and the fact that the sensors used for parameter measurements are not always reliable. In such scenarios, the Gaussian assumption will not hold and the accuracy of the Kalman filter will be affected or even lead to dispersion of the results\cite{ref4,ref5,ref6,ref7,ref8}.

To solve the above problems, a large number of research results have been published. Examples are the Huber-based Kalman filter (HKF) \cite{ref9} and the maximum correntropy Kalman filter (MCKF) \cite{ref10,ref11,ref12,ref13}, but both of them are derived by minimizing or maximizing the cost function without considering the nature of heavy-tailed noise \cite{ref14,ref15}, which can lead to deteriorating estimation results.

Student's t filter dealt with the problem, The key idea of it is to approximate the posterior probability density function (PDF) by a Student’s t PDF \cite{ref16,ref17,ref18}, but the Student's t filter used moment matching approach to preserve the heavy-tailed properties, this will only capture the first few moments and can lead to the lost of higher order moments \cite{ref19}.

TKF solved the problem mentioned in the previous paragraph, it approximated the one-step prediction PDF and the likelihood PDF as the student's t-distribution with different parameters of degrees of freedom and models the unknown scale matrix of the one-step prediction PDF as an inverse Wishart distribution \cite{ref20}, and TKF has a stable and excellent filtering effect when the measurement noise has a heavy-tailed nature.

The above filters solved the problem of accuracy under different noise conditions, and they all essentially adjust the proportion of the predicted and measurement vectors in the filtering result to make it credible. However, TKF paid more attention to the heavy-tailed part of the noise, so TKF introduced constant parameters to the algorithm to assist in the adjustment of the heavy-tailed part of the noise to obtain better results, but when adjusting the Gaussian part, the adjustment result can be less accurate due to the influence of the introduced constant parameters. In practical applications and experiments, the proportion of Gaussian noise in the measurement noise remains high, so the adjustment method of TKF needs to be further optimized.

This brief analyzed how the constant parameter introduced to TKF affects the proportion adjustment result, and proposed TGKF, TGKF retains the accuracy of TKF under heavy-tailed noise conditions, meanwhile, TGKF has an optimized filtering effect when facing Gaussian noise. Under the simulation setup conditions, the results show that TGKF has a smaller RMSE than TKF and KF.

\section{Brief Review of Kalman Filter}
Assume that a discrete-time linear dynamic system is modeled as follows:
\begin{align}
    &\textbf{\textit{x}}_\textit{k} = \textbf{\textit{f}}_{\textit{k}-1}\textbf{\textit{x}}_{\textit{k}-1}+\textbf{\textit{w}}_{\textit{k}-1},\\
    &\textbf{\textit{z}}_\textit{k} = \textbf{\textit{h}}_\textit{k}\textbf{\textit{x}}_\textit{k}+\textbf{\textit{v}}_\textit{k},
\end{align}

where $\textit{k}$ is the discrete time series number, $\textbf{\textit{x}}_\textit{k}$ is the n-dimensional state vector, $\textbf{\textit{z}}_k$ is the m-dimensional measurement vector, $\textbf{\textit{f}}$  is the n$\times$n state transfer matrix, $\textbf{\textit{h}}$  is the m$\times$n state observation matrix, $\textbf{\textit{w}}_\textit{k}$  is the n-dimensional state noise vector with zero mean and nominal covariance matrix $\textbf{\textit{q}}_\textit{k}$, $\textbf{\textit{v}}_\textit{k}$ is the m-dimensional measurement noise vector with zero mean and covariance matrix $\textbf{\textit{r}}_\textit{k}$. Due to the existence of errors and outliers in the measurement process, $\textbf{\textit{q}}_\textit{k}$ and $\textbf{\textit{r}}_\textit{k}$ are not completely accurate.

Under the circumstance that the state-space model and measurement vector are given, the Kalman filter is a commonly used method for inferring the state vector ${\textbf{\textit{x}}_\textit{k}}$, and the iterative update process is shown below:

\begin{equation}
\begin{aligned}
&\textbf{\textit{\^{x}}}_{\textit{k}|\textit{k}-1} = \textbf{\textit{f}}_{\textit{k}-1}\textbf{\textit{\^{x}}}_{\textit{k}-1|\textit{k}-1},\\
    &\textbf{\textit{p}}_{\textit{k}|\textit{k}-1} = \textbf{\textit{f}}_{\textit{k}-1}\textbf{\textit{p}}_{\textit{k}-1|\textit{k}-1}{\textbf{\textit{f}}^T_{\textit{k}-1}}+{\textbf{\textit{q}}_{\textit{k}-1}},\\
    &\textbf{\textit{k}}_\textit{k} = \textbf{\textit{p}}_{\textit{k}|\textit{k}-1}\textbf{\textit{h}}^T_\textit{k}(\textbf{\textit{h}}_\textit{k}\textbf{\textit{p}}_{\textit{k}|\textit{k}-1}\textbf{\textit{h}}^T_\textit{k}+\textbf{\textit{r}}_\textit{k})^{-1},\\
    &\textbf{\textit{\^{x}}}_{\textit{k}|\textit{k}} = \textbf{\textit{\^{x}}}_{\textit{k}|\textit{k}-1}+\textbf{\textit{k}}_\textit{k}(\textbf{\textit{z}}_\textit{k}-\textbf{\textit{h}}_\textit{k}\textbf{\textit{\^{x}}}_{\textit{k}|\textit{k}-1}),\\
    &\textbf{\textit{p}}_{\textit{k}|\textit{k}} = \textbf{\textit{p}}_{\textit{k}|\textit{k}-1}-\textbf{\textit{k}}_\textit{k}\textbf{\textit{h}}_\textit{k}\textbf{\textit{p}}_{\textit{k}|\textit{k}-1},
\end{aligned}
\end{equation}

where $()^T$ represents the matrix transpose operation, $\textbf{\textit{\^{x}}}_{\textit{k}|\textit{k}-1}$ and $\textbf{\textit{p}}_{\textit{k}|\textit{k}-1}$ are the prior estimated state vector of step $\textit{k}$ with its corresponding error covariance matrix, $\textbf{\textit{\^{x}}}_{\textit{k}|\textit{k}}$ and $\textbf{\textit{p}}_{\textit{k}|\textit{k}}$ are the posterior estimated state vector of step $\textit{k}$ with its corresponding error covariance matrix, and $\textbf{\textit{k}}_\textit{k}$ is the Kalman gain at step $\textit{k}$.

The essence of KF is to process the posterior estimated state vector at step $\textit{k}-1$ by the state transfer matrix to become the prior estimated state vector at step $\textit{k}$, and combine this with the measurement vector at step $\textit{k}$, then adjust the proportion of both vectors by Kalman gain to finally obtain a reliable posterior estimated state at step $\textit{k}$, to clearly explain the working principle of KF, we can firstly make the following derivation($\textbf{\textit{I}}$ stands for unit matrix):

\begin{equation}
\begin{aligned}
\textbf{\textit{\^{x}}}_{\textit{k}|\textit{k}} &= \textbf{\textit{\^{x}}}_{\textit{k}|\textit{k}-1}+\textbf{\textit{k}}_\textit{k}(\textbf{\textit{z}}_\textit{k}-\textbf{\textit{h}}_\textit{k}\textbf{\textit{\^{x}}}_{\textit{k}|\textit{k}-1})\\
&= \textbf{\textit{\^{x}}}_{\textit{k}|\textit{k}-1}\textbf{\textit{k}}_\textit{k}\textbf{\textit{z}}_\textit{k}-\textbf{\textit{k}}_\textit{k}\textbf{\textit{h}}_\textit{k}\textbf{\textit{\^{x}}}_{\textit{k}|\textit{k}-1}\\
&=(\textbf{\textit{I}}-\textbf{\textit{k}}_\textit{k}\textbf{\textit{h}}_\textit{k})\textbf{\textit{\^{x}}}_{\textit{k}|\textit{k}-1}+\textbf{\textit{k}}_\textit{k}\textbf{\textit{z}}_\textit{k},
\end{aligned}
\end{equation}

Substitute the expression of  $\textbf{\textit{k}}_\textit{k}$ into the expression of $\textbf{\textit{p}}_{\textit{k}|\textit{k}}$, and by using the matrix inverse formula we have:
\begin{align}
\textbf{\textit{p}}_{\textit{k}|\textit{k}}\textbf{\textit{h}}^T_\textit{k} = \textbf{\textit{k}}_\textit{k}\textbf{\textit{r}}_{\textit{k}},
\end{align}

Based on equations (4) and (5), the state estimation equations of KF can be simplified and expressed in the following form by applying the matrix inverse formula again \cite{ref21}:
\begin{align}
    &\textbf{\textit{\^{x}}}_{\textit{k}|\textit{k}} = \textbf{\textit{A}}_\textit{p}\textbf{\textit{\^{x}}}_{\textit{k}|\textit{k}-1}+\textbf{\textit{A}}_\textit{r}\textbf{\textit{z}}_\textit{k},\\
&\textbf{\textit{A}}_\textit{p} = [(\textbf{\textit{p}}_{\textit{k}|\textit{k}-1})^{-1}+\textbf{\textit{h}}^T_\textit{k}(\textbf{\textit{r}}_{\textit{k}})^{-1}\textbf{\textit{h}}_\textit{k}]^{-1}(\textbf{\textit{p}}_{\textit{k}|\textit{k}-1})^{-1},\\
&\textbf{\textit{A}}_\textit{r} = [(\textbf{\textit{p}}_{\textit{k}|\textit{k}-1})^{-1}+\textbf{\textit{h}}^T_\textit{k}(\textbf{\textit{r}}_{\textit{k}})^{-1}\textbf{\textit{h}}_\textit{k}]^{-1}\textbf{\textit{h}}^T_\textit{k}(\textbf{\textit{r}}_{\textit{k}})^{-1},
\end{align}

From equations (6)-(8), we can obtain that the posterior estimated state at step $\textit{k}$, $\textbf{\textit{\^{x}}}_{\textit{k}|\textit{k}}$ is the linear combination of prior estimated state $\textbf{\textit{\^{x}}}_{\textit{k}|\textit{k}-1}$ and measurement state $\textbf{\textit{z}}_\textit{k}$. According to the different state noises and measurement noises appearing in the real situation, the matrix $\textbf{\textit{A}}_\textit{p}$ and $\textbf{\textit{A}}_\textit{r}$ can adjust the weights of $\textbf{\textit{\^{x}}}_{\textit{k}|\textit{k}-1}$ and $\textbf{\textit{z}}_\textit{k}$ in the final estimation with a targeted way, where the parameters ${\textbf{\textit{p}}}_{\textit{k}|\textit{k}-1}$ and ${\textbf{\textit{r}}}_{\textit{k}}$ are quite crucial to the adjustment of the weights, and ${\textbf{\textit{p}}}_{\textit{k}|\textit{k}-1}$ are determined by $\textbf{\textit{q}}_{\textit{k}-1}$. The following figures show actually how KF adjusts the value of $\textbf{\textit{A}}_\textit{p}$ and $\textbf{\textit{A}}_\textit{r}$ when Gaussian and impulsive noises come.

\begin{figure}[!h]
\centering
\includegraphics[width=3in]{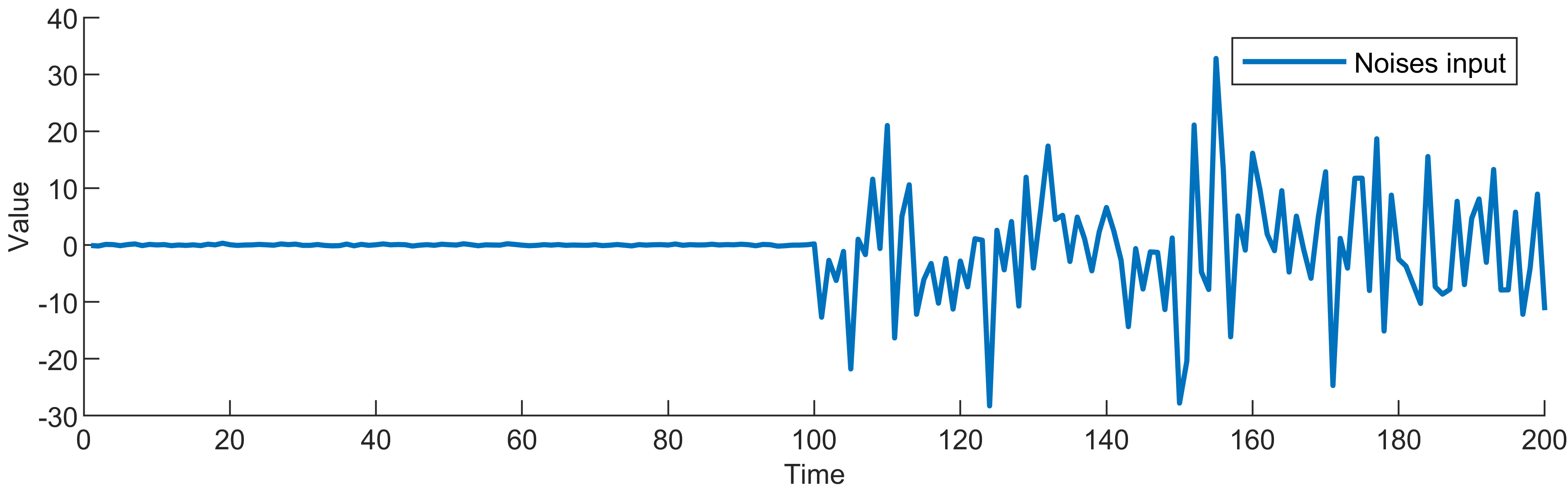}
\caption{Noises input.}
\label{fig_1}
\end{figure}

To demonstrate more intuitively how KF works, a one-dimensional motion model $\textit{x} = \textit{x}+1$ is used, so that $\textbf{\textit{A}}_\textit{p}$ and $\textbf{\textit{A}}_\textit{r}$ can be expressed by numbers, 200 noises are inputted for simulation, where the first 100 steps of the inputted noises are Gaussian and 101-200 steps are impulsive.

\begin{figure}[!h]
\centering
\includegraphics[width=3in]{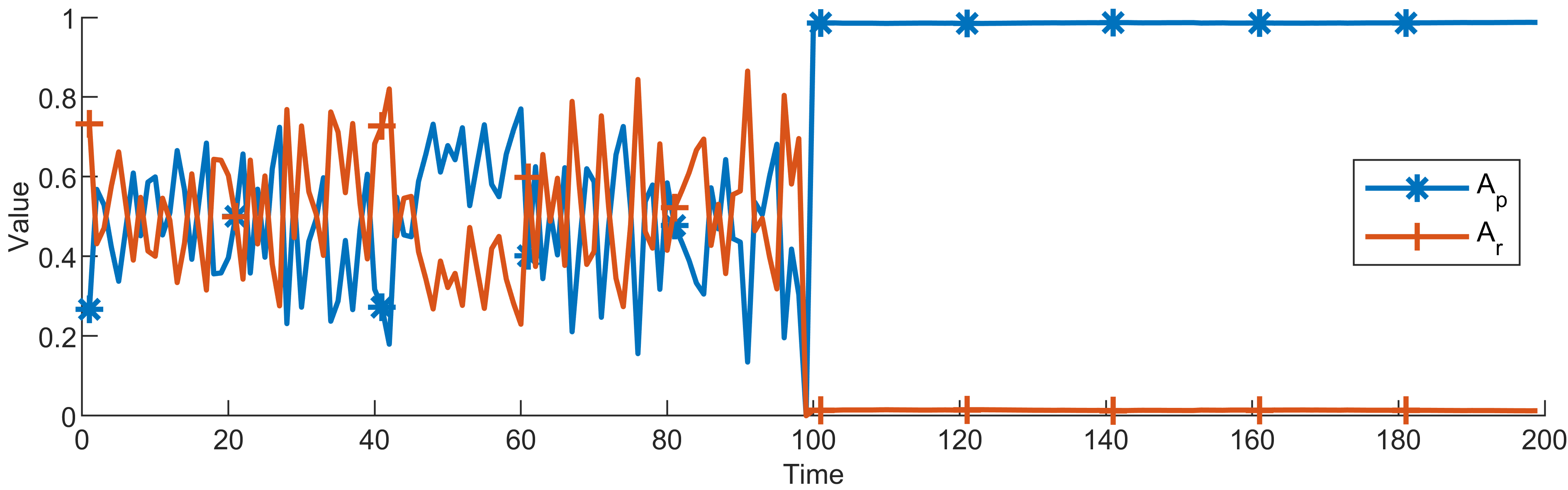}
\caption{Values of $\textbf{\textit{A}}_\textit{p}$ and $\textbf{\textit{A}}_\textit{r}$.}
\label{fig_2}
\end{figure}

As can be seen from the first 100 data in Fig. 2, the value of $\textbf{\textit{A}}_\textit{p}$ and $\textbf{\textit{A}}_\textit{r}$ fluctuate up and down around 0.5, when $\textbf{\textit{A}}_\textit{P}$ rises, $\textbf{\textit{A}}_\textit{R}$ falls, and vice versa. The reason for this phenomenon is that the process noises and measurement noises exist at the same time, and the confidence level of measurement value and estimated value keeps changing, and the confidence level of one side rises will inevitably lead to the decline of the other side. At the same time, due to the vast majority of the process noises being also Gaussian, so at most of the time gap of the confidence level between the measurement value and estimated value is not too large, resulting in $\textbf{\textit{A}}_\textit{p}$ and $\textbf{\textit{A}}_\textit{r}$ fluctuations in the vicinity of 0.5. The reason for the large fluctuations at a few nodes is the presence of impulsive process noises.

And from the last 100 data in Fig. 2, the values of $\textbf{\textit{A}}_\textit{p}$ are close to 1, in the meanwhile, the values of $\textbf{\textit{A}}_\textit{r}$ are close to 0 and they vary very little. The reason for this phenomenon is that when the inputted measurement noises are impulsive, which means measurement values are no longer trustworthy, KF will adjust the confidence level of the estimated value ($\textbf{\textit{A}}_\textit{p}$) to a very high level to ensure that the result is credible.

KF has a high degree of credibility when filtering models with Gaussian noise and linear state space. However, in practical experiments, heavy-tailed processes and measurement noise violate the Gaussian assumption of classical Kalman filtering, making it a sub-optimal choice in such situations.

\section{Explanation of This Work}

Similar to what was mentioned in Section 2, we also have a simplified expression of the existing student's t based Kalman filter as follows \cite{ref20}:
\begin{align}
    &\textbf{\textit{\^{X}}}^{\textit{i}+1}_{\textit{k}|\textit{k}} = \textbf{\textit{C}}_\textit{P}\textbf{\textit{\^{X}}}^{\textit{i}+1}_{\textit{k}|\textit{k}-1}+\textbf{\textit{C}}_\textit{R}\textbf{\textit{Z}}_\textit{k},\\
&\textbf{\textit{C}}_\textit{P} = [(\widetilde{\textbf{\textit{P}}}^{\textit{i}+1}_{\textit{k}|\textit{k}-1})^{-1}+\textbf{\textit{H}}^T_\textit{k}(\widetilde{\textbf{\textit{R}}}^{\textit{i}+1}_{\textit{k}})^{-1}\textbf{\textit{H}}_\textit{k}]^{-1}(\widetilde{\textbf{\textit{P}}}^{\textit{i}+1}_{\textit{k}|\textit{k}-1})^{-1},\\
&\textbf{\textit{C}}_\textit{R} = [(\widetilde{\textbf{\textit{P}}}^{\textit{i}+1}_{\textit{k}|\textit{k}-1})^{-1}+\textbf{\textit{H}}^T_\textit{k}(\widetilde{\textbf{\textit{R}}}^{\textit{i}+1}_{\textit{k}})^{-1}\textbf{\textit{H}}_\textit{k}]^{-1}\textbf{\textit{H}}^T_\textit{k}(\widetilde{\textbf{\textit{R}}}^{\textit{i}+1}_{\textit{k}})^{-1},
\end{align}

The equations related to $\textbf{\textit{R}}_{\textit{k}}$ in TKF are listed below for further analysis:
\begin{align}
&\widetilde{\textbf{\textit{R}}}^{\textit{i}+1}_{\textit{k}} = \textbf{\textit{R}}_{\textit{k}}/[\textbf{\textit{E}}^{\textit{i}+1}(\textit{$\lambda$}_\textit{k})],\\
&\textbf{\textit{E}}^{\textit{i}+1}(\textit{$\lambda$}_\textit{k}) = (\textit{m}+\textit{v})/[\textit{v}+\textit{tr}(\textbf{\textit{E}}^{\textit{i}+1}_\textit{k}\textbf{\textit{R}}^{-1}_\textit{k})],\\
&\textbf{\textit{E}}^{\textit{i}+1}_\textit{k} = (\textbf{\textit{Z}}_\textit{k}-\textbf{\textit{H}}_\textit{k}\textbf{\textit{\^{X}}}^{\textit{i}+1}_{\textit{k}|\textit{k}})(\textbf{\textit{Z}}_\textit{k}-\textbf{\textit{H}}_\textit{k}\textbf{\textit{\^{X}}}^{\textit{i}+1}_{\textit{k}|\textit{k}})^T+\textbf{\textit{H}}_\textit{k}\textbf{\textit{P}}^{\textit{i}+1}_{\textit{k}|\textit{k}}\textbf{\textit{H}}^T_\textit{k},
\end{align} 
In the following derivation: let $\textbf{\textit{R}}_\textit{k}(\textit{S})$ represents the covariance matrix of the Gaussian part in the measurement noise, $\textbf{\textit{R}}_\textit{k}(\textit{N})$ represents the covariance matrix of the measurement noise, and $\textbf{\textit{R}}_\textit{k}(\textit{B})$ represents the covariance matrix of the impulsive part in the measurement noise, it is easy to obtain:
\begin{align}
    \textbf{\textit{R}}_\textit{k}(\textit{B})>\textbf{\textit{R}}_\textit{k}(\textit{N})>\textbf{\textit{R}}_\textit{k}(\textit{S}),
\end{align}

When the impulse noise appears, by Equation (14) and (12), the value of $\textbf{\textit{Z}}_\textit{k}-\textbf{\textit{H}}_\textit{k}\textbf{\textit{\^{X}}}^{\textit{i}+1}_{\textit{k}|\textit{k}}$ will increase substantially, which will lead to a large increase in $\widetilde{\textbf{\textit{R}}}^{\textit{i}+1}_{\textit{k}}$, and then by Equation (11), this will eventually lead to a large reduction in the proportion of the measurement vector in the final estimation result, in this situation, TKF has a considerable effect.

When Gaussian noise presents, Similar to those described in the previous paragraph, the value of $\textbf{\textit{Z}}_\textit{k}-\textbf{\textit{H}}_\textit{k}\textbf{\textit{\^{X}}}^{\textit{i}+1}_{\textit{k}|\textit{k}}$ will decrease substantially. The adjustment variable tries to maximize $\textbf{\textit{E}}^{\textit{i}+1}(\lambda_\textit{k})$ to maximize the proportion of the measurement vector in the final estimation result, but due to the presence of the constants($\textit{m}\text{ }and\text{ }\textit{v}$) in Equation (13), no matter how small $\textbf{\textit{E}}^{\textit{i}+1}_\textit{k}$ is, $\textbf{\textit{E}}^{\textit{i}+1}(\lambda_\textit{k})$ could only get infinitely close to the adjustment limit $({\textit{m}+\textit{v}})/{\textit{v}}$. This indicates that TKF is not sufficiently adjusting the weight of the measurement vector. At the same time, the impulse noises included in measurement noises will always ensure that inequality (15) holds.

The above reasoning suggests that when Gaussian noise is presented, the existing algorithm is not accurate enough to adjust the weight of the measurement vector in the final estimation result by adjusting a single parameter $\textbf{\textit{E}}^{\textit{i}+1}(\lambda_\textit{k})$, this provides an idea to optimize the prior work, that is to replace $\textbf{\textit{R}}_\textit{k}(\textit{N})$ in the existing algorithm with $\textbf{\textit{R}}_\textit{k}(\textit{S})$, this can break the adjustment limit and further increase the weight of the measurement vector in the final estimation result.

However, directly replacing $\textbf{\textit{R}}_\textit{k}(\textit{N})$ with $\textbf{\textit{R}}_\textit{k}(\textit{S})$ will affect the accuracy of the filtering results when there is a high percentage of impulse noise in measurement noise (approximately when greater than 55$\%$). To make the algorithm robust, an adjustment factor for $\textbf{\textit{R}}_\textit{k}(\textit{S})$ is proposed, we named it $\textit{TG}$, it adapts with the share of the two Gaussian distributions in the GMM decomposition results.

\begin{align}
    \textit{TG} = 4\sqrt{det[\textbf{\textit{R}}_\textit{k}(\textit{B})/\textbf{\textit{R}}_\textit{k}(\textit{S})]}/[3+\exp(10\textit{P})],
\end{align}
where $\textit{P}$  represents the proportion of Gaussian noise in measurement noise. As adjusted by $\textit{TG}$, TGKF can cope with up to the situation that 70$\%$ of measurement noises are impulsive without being inferior, in the meanwhile, the accuracy will not be affected in the case of high Gaussian percentage in measurement noise. The simulation result of Gaussian percentage is shown in the simulation part.

GMM is generally used to model statistical models of probability distributions, if the data set is generated by multiple different Gaussian distributions, with each having a certain weight, indicating the probability of the corresponding pattern appearing. The probability distribution of the Gaussian mixture model is:

\begin{align}
    P(\textit{u}|\theta) = \Sigma_{\textit{k}=1}^\textit{K}\alpha_\textit{k}\Phi(\textit{u}|\theta_\textit{k}),
\end{align}

Where $\textit{u}$ is the v-dimensional observed data, $\textit{K}$ is the number of sub-Gaussian models in the mixture model, $\textit{k} = 1,2,3...,\textit{K}$, $\alpha_\textit{k}$ is the probability that the observation belongs to the $\textit{k}$th sub-model, $\alpha_\textit{k} \geq 0$, $\Sigma_{\textit{k}=1}^\textit{K}\alpha_\textit{k} = 1$ and $\Phi(\textit{u}|\theta_\textit{k})$ is the Gaussian distribution density function of the $\textit{k}$th sub-model, $\theta_\textit{k} = (\mu_\textit{k}, \sigma_\textit{k}^2)$, $\mu_\textit{k}$ is the mean of the Gaussian distribution of the $\textit{k}$th sub-model and $\sigma_\textit{k}^2$  is the covariance matrix of the Gaussian distribution for the $\textit{k}$th sub-model, $\Phi(\textit{u}|\theta_\textit{k}) = (1/{\sqrt{(2\pi)^\textit{v}|\Sigma_\textit{k}|}}){exp}[-\frac{1}{2}(\textit{u}-\mu_\textit{k})^\textit{T}\Sigma_\textit{k}^{-1}(\textit{u}-\mu_\textit{k})]$.

Single Gaussian distributions from mixed Gaussian distributions can be separated by learning the parameters. Here we only consider the case where GMM decomposes two Gaussian distributions, noises that are mixed by two Gaussian distributions are very common in practical application scenarios \cite{ref22}, by adjusting the covariance matrix of the two Gaussian distributions and the ratio of them in the mixture, this is sufficient to fit most of the types of noises. GMM works well with mixed Gaussian noises, when other noises present, using GMM for approximation, the results are also reliable. In this situation, GMM will be used for the estimation of $\textbf{\textit{R}}_\textit{k}(\textit{S})$ and $\textbf{\textit{R}}_\textit{k}(\textit{B})$.

The proposed filter is expressed by the pseudocode below \cite{ref20}:
\begin{algorithm}
\setstretch{1.2}
\caption{One time step of the covariance adaptive student's t based Kalman filter} \label{alg:alg1}
\begin{algorithmic}
\STATE 
 $\textbf{GMM is applied before the algorithm to get}\text{ } \textbf{\textit{R}}_{\textit{k}}(\textit{S}),\text{ }\textbf{\textit{R}}_{\textit{k}}(\textit{B}).$
    
    \textbf{Inputs}: $\textbf{\textit{\^{X}}}_{\textit{k}-1|\textit{k}-1}$, $\textbf{\textit{P}}_{\textit{k}-1|\textit{k}-1}$, $\textbf{\textit{F}}_{\textit{k}-1}$, $\textbf{\textit{H}}_{\textit{k}}$, $\textbf{\textit{Z}}_{\textit{k}}$, $\textbf{\textit{Q}}_{\textit{k}-1}$, ${\textbf{\textit{R}}}_{\textit{k}}(\textit{S})$, ${\textbf{\textit{R}}}_{\textit{k}}(\textit{B})$, $\textit{m}$, $\textit{n}$, $\omega$, $\upsilon$, $\tau$, $\textit{N}$
    
    \textbf{Time update}:
    
    1.  $\textbf{\textit{\^{X}}}_{\textit{k}|\textit{k}-1}$ = $\textbf{\textit{F}}_{\textit{k}-1}$$\textbf{\textit{\^{X}}}_{\textit{k}-1|\textit{k}-1}$
    
    2. $\textbf{\textit{P}}_{\textit{k}|\textit{k}-1}$ = $\textbf{\textit{F}}_{\textit{k}-1}$$\textbf{\textit{P}}_{\textit{k}-1|\textit{k}-1}$$\textbf{\textit{F}}^{\textit{T}}_{\textit{k}-1}$ + $\textbf{\textit{Q}}_{\textit{k}-1}$

   \textbf{Measurement update}:
    
    3. Initialization: $\textit{u}_\textit{k} = \textit{n}+\tau+1$, $\textbf{\textit{U}}_\textit{k} = \tau\textbf{\textit{P}}_{\textit{k}|\textit{k}-1}$, $\textbf{\textit{\^{X}}}_{\textit{k}|\textit{k}}^{(0)} = \textbf{\textit{\^{X}}}_{\textit{k}|\textit{k}-1}$, $\textbf{\textit{P}}_{\textit{k}|\textit{k}}^{(0)} = \textbf{\textit{P}}_{\textit{k}|\textit{k}-1}$, $E^{(0)}[\Sigma_\textit{k}^{-1}] = (\textit{u}_\textit{k}-\textit{n}-1)\textbf{\textit{U}}_\textit{k}^{-1}$, $\textbf{\textit{R}}_\textit{k}(\textit{I}) = \textit{TG}*\textbf{\textit{R}}_\textit{k}(\textit{S})$

     \textbf{for} \textit{i} = 0 : \textit{N }- 1

     4. $\textbf{\textit{D}}_\textit{k}^{(\textit{i})} = \textbf{\textit{P}}_{\textit{k}|\textit{k}}^{(i)}+(\textbf{\textit{\^{X}}}_{\textit{k}|\textit{k}}^{(i)}-\textbf{\textit{\^{X}}}_{\textit{k}|\textit{k}-1})(\textbf{\textit{\^{X}}}_{\textit{k}|\textit{k}}^{(i)}-\textbf{\textit{\^{X}}}_{\textit{k}|\textit{k}-1})^{\textit{T}}$

     5. $\alpha_\textit{k}^{\textit{i}+1} = 0.5(\textit{n}+\omega)$, $\beta_\textit{k}^{\textit{i}+1} = 0.5\{\omega+tr(\textbf{\textit{D}}_\textit{k}^{(\textit{i})})E^{(\textit{i})}[\Sigma_\textit{k}^{-1}])\}$, $E^{({\textit{i}+1})}[\xi_\textit{k}] = \alpha_\textit{k}^{\textit{i}+1}/\beta_\textit{k}^{\textit{i}+1}$
     
     6. $\textbf{\textit{E}}_\textit{k}^{(\textit{i})} = (\textbf{\textit{Z}}_\textit{k}-\textbf{\textit{H}}_\textit{k}\textbf{\textit{\^{X}}}_{\textit{k}|\textit{k}-1}^{(\textit{i})})(\textbf{\textit{Z}}_\textit{k}-\textbf{\textit{H}}_\textit{k}\textbf{\textit{\^{X}}}_{\textit{k}|\textit{k}-1}^{(\textit{i})})^\textit{T}+\textbf{\textit{H}}_\textit{k}\textbf{\textit{P}}_{\textit{k}|\textit{k}}^{(i)}\textbf{\textit{H}}_\textit{k}^\textit{T}$

     7. $\gamma_\textit{k}^{\textit{i}+1} = 0.5(\textit{m}+\upsilon)$, $\delta_\textit{k}^{\textit{i}+1} = 0.5\{\upsilon+tr(\textbf{\textit{E}}_\textit{k}^{(\textit{i})}\textbf{\textit{R}}_\textit{k}(\textit{I})^{-1})\}$, $\textbf{\textit{E}}^{(\textit{i}+1)}[\lambda_\textit{k}] = \gamma_\textit{k}^{\textit{i}+1}/\delta_\textit{k}^{\textit{i}+1}$

     8. $\textbf{\textit{\^{u}}}_\textit{k}^{(\textit{i}+1)} = \textbf{\textit{u}}_{\textit{k}}+1$, $\textbf{\textit{\^{U}}}_\textit{k}^{(\textit{i}+1)} = \textbf{\textit{U}}_\textit{k}+\textbf{\textit{E}}^{(\textit{i}+1)}[\xi_\textit{k}]\textbf{\textit{D}}_\textit{k}^{(\textit{i})}$, $E^{(\textit{i}+1)}[\Sigma_\textit{k}^{-1}] = (\textbf{\textit{\^{u}}}_\textit{k}^{(\textit{i}+1)}-\textit{n}-1)(\textbf{\textit{\^{U}}}_\textit{k}^{(\textit{i}+1)})^{-1}$

     9. $\widetilde{\textbf{\textit{R}}}^{(\textit{i}+1)}_{\textit{k}} = \textbf{\textit{R}}_\textit{k}(\textit{I})/\textbf{\textit{E}}^{(\textit{i}+1)}[\lambda_\textit{k}]$, 
     
     $\widetilde{\textbf{\textit{P}}}^{(\textit{i}+1)}_{\textit{k}|\textit{k}-1} = \{\textbf{\textit{E}}^{(\textit{i}+1)}[\Sigma_\textit{k}^{-1}]\}^{-1}/E^{(\textit{i}+1)}[\xi_\textit{k}]$,

     10. $\textbf{\textit{K}}^{(\textit{i}+1)}_{\textit{k}} = \widetilde{\textbf{\textit{P}}}^{(\textit{i}+1)}_{\textit{k}|\textit{k}-1}\textbf{\textit{H}}^\textit{T}_\textit{k}(\textbf{\textit{H}}_\textit{k}\widetilde{\textbf{\textit{P}}}^{(\textit{i}+1)}_{\textit{k}|\textit{k}-1}\textbf{\textit{H}}^\textit{T}_\textit{k}+\widetilde{\textbf{\textit{R}}}^{(\textit{i}+1)}_{\textit{k}})^{-1}$

     11.$\textbf{\textit{\^{X}}}_{\textit{k}|\textit{k}}^{(\textit{i}+1)} = \textbf{\textit{\^{X}}}_{\textit{k}|\textit{k}-1}+\textbf{\textit{K}}^{(\textit{i}+1)}_{\textit{k}}(\textbf{\textit{Z}}_\textit{k}-\textbf{\textit{H}}_\textit{k}\textbf{\textit{\^{X}}}_{\textit{k}|\textit{k}-1})$

     12.$\textbf{\textit{P}}_{\textit{k}|\textit{k}}^{(\textit{i}+1)} = \widetilde{\textbf{\textit{P}}}^{(\textit{i}+1)}_{\textit{k}|\textit{k}-1}-\textbf{\textit{K}}^{(\textit{i}+1)}_{\textit{k}}\textbf{\textit{H}}_\textit{k}\widetilde{\textbf{\textit{P}}}^{(\textit{i}+1)}_{\textit{k}|\textit{k}-1}$

     \textbf{end for}

     13.$\textbf{\textit{\^{X}}}_{\textit{k}|\textit{k}} = \textbf{\textit{\^{X}}}_{\textit{k}|\textit{k}}^{(\textit{N})}$, $\textbf{\textit{P}}_{\textit{k}|\textit{k}} = \textbf{\textit{P}}_{\textit{k}|\textit{k}}^{(\textit{N})}$

      \textbf{Outputs:} $\textbf{\textit{\^{X}}}_{\textit{k}|\textit{k}}$ and $\textbf{\textit{P}}_{\textit{k}|\textit{k}}$
      
\end{algorithmic}
\label{alg1}
\end{algorithm}

\section{Simulation}

From Equations (9)-(14), we can know that $\widetilde{\textbf{\textit{R}}}^{\textit{i}+1}_{\textit{k}}$ and $\textbf{\textit{E}}^{(\textit{i}+1)}[\lambda_\textit{k}]$ can both adjust the size of $\textbf{\textit{C}}_\textit{R}$, when the size of $\textbf{\textit{C}}_\textit{R}$ increases, the proportion of the measurement vector in the final result will further increase, which makes the filtering result more trustworthy under Gaussian noise conditions.

The aforementioned TGKF, KF, and TKF, three filters are simulated and compared in this section.
\textbf{\subsection{Pre-Simulation}}

In the simulation work, a two-dimensional uniform motion model is considered with the following equations of state and trajectory equations:

\begin{equation}
\textbf{\textit{X}}_\textit{k} = \begin{bmatrix} 1 & 0 & \Delta T & 0 \\ 
0 &  1 & 0 & \Delta T \\ 0 & 0 & 1 & 0\\ 0 & 0 & 0 & 1\end{bmatrix}\textbf{\textit{X}}_{\textit{k}-1}+\begin{bmatrix} \textbf{\textit{Q}}_{1;\textit{k}-1}\\\textbf{\textit{Q}}_{2;\textit{k}-1}\\\textbf{\textit{Q}}_{3;\textit{k}-1}\\\textbf{\textit{Q}}_{4;\textit{k}-1}  \end{bmatrix},
\end{equation}

\begin{equation}
\textbf{\textit{Y}}_\textit{k} = \begin{bmatrix} 1 & 0 & 0 & 0 \\ 
0 & 0 & 0 & 1\end{bmatrix}\textbf{\textit{X}}_{\textit{k}}+\textbf{\textit{V}}_\textit{k},
\end{equation}

While $\textbf{\textit{X}}_\textit{k} = \begin{bmatrix} \textit{X}_\textit{k} & \textit{Y}_\textit{k} & \dot{\textit{X}}_\textit{k} & \dot{\textit{Y}}_\textit{k}\end{bmatrix}$, $\textit{X}_\textit{k}$, $\textit{Y}_\textit{k}$, $\dot{\textit{X}}_\textit{k}$, $\dot{\textit{Y}}_\textit{k}$ represent the Cartesian coordinate system positions and the corresponding velocities. $\textit{Q}_{n;\textit{k}-1}$ represents the $\textit{n}$th element of the state noises at step $\textit{k}-1$, while $\textbf{\textit{V}}_\textit{k}$ represents the measurement noise at step  $\textit{k}$. The covariance matrix of the state noise is set as follows:

\begin{equation}
\textbf{\textit{Q}}_\textit{k} = \begin{bmatrix} \frac{\Delta\textit{T}^3}3 & 0 & \frac{\Delta\textit{T}^2}3 & 0 \\ 
0 &  \frac{\Delta\textit{T}^3}3 & 0 & \frac{\Delta\textit{T}^2}3 \\ \frac{\Delta\textit{T}^3}2 & 0 & \Delta\textit{T}^2 & 0\\ 0 & \frac{\Delta\textit{T}^2}2 & 0 & \Delta\textit{T}^2\end{bmatrix},
\end{equation}

The time interval is $\Delta\textit{T} = 1\textit{second}$, the state initial value $\textbf{\textit{X}}_0 = \begin{bmatrix} 0 & 0 & 15 & 15\end{bmatrix}$, $\textbf{\textit{\^{X}}}_{0|0} = \textbf{\textit{F}}_{\textit{k}-1}\textbf{\textit{\^{X}}}_{\textit{k}-1|\textit{k}-1} = \begin{bmatrix} 15 & 15 & 15 & 15\end{bmatrix}$, it simulates a motion model in which the object starts from position $\begin{bmatrix} 0 & 0\end{bmatrix}$ and travels at a constant speed of 15 both in the x-coordinate direction and the y-coordinate direction. $\textbf{\textit{P}}_{0|0} = \textbf{\textit{I}}_\textit{n}$,  where $\textbf{\textit{I}}_\textit{n}$ is a unit matrix of order n. The degree of freedom parameters of the student's t distribution are $\omega = \nu = 5$, and the tuning parameter $\tau = 5$.

Considering the measurement noises follow mixed Gaussian distributions: (w.p. means with probability)

\begin{align}
{\textbf{\textit{V}}_\textit{k}} = 
\begin{cases}
\textit{N}(\textbf{0},0.1\textbf{\textit{I}}_2),&{\text{w.p. 0.90}},\\ 
{\textit{N}(\textbf{0},10\textbf{\textit{I}}_2),}&{\text{w.p. 0.10}}, 
\end{cases}
\end{align}

To compare the performance of KF, TKF, and TGKF, we compared the results of localization trajectory simulations and the root mean square errors (RMSEs) of these filters.

\begin{align}
    \textit{RMSE} = \sqrt{\frac{1}{\textit{M}}\Sigma_{\textit{s}=1}^\textit{M}((\textbf{\textit{X}}_\textit{k}^\textit{s}-\textbf{\textit{\^X}}_\textit{k}^\textit{s})+(\textbf{\textit{Y}}_\textit{k}^\textit{s}-\textbf{\textit{\^Y}}_\textit{k}^\textit{s}))},
\end{align}

All filters are coded with MATLAB R2022b and simulations are run on a computer with Intel Core i5-12490F CPU.

\textbf{\subsection{Simulation Results}}
The simulation results are plotted after averaging 500 times of operations on the state variable $\textbf{\textit{X}}$.

The simulation results of the true and estimated trajectories obtained from KF, TKF, and TGKF are shown in Fig. 3 (only some data intervals are intercepted for clear presentation), and the RMSEs of positions obtained from KF, TKF, and TGKF are respectively shown in Fig. 4:

\begin{figure}[!h]
\centering
\includegraphics[width=3in]{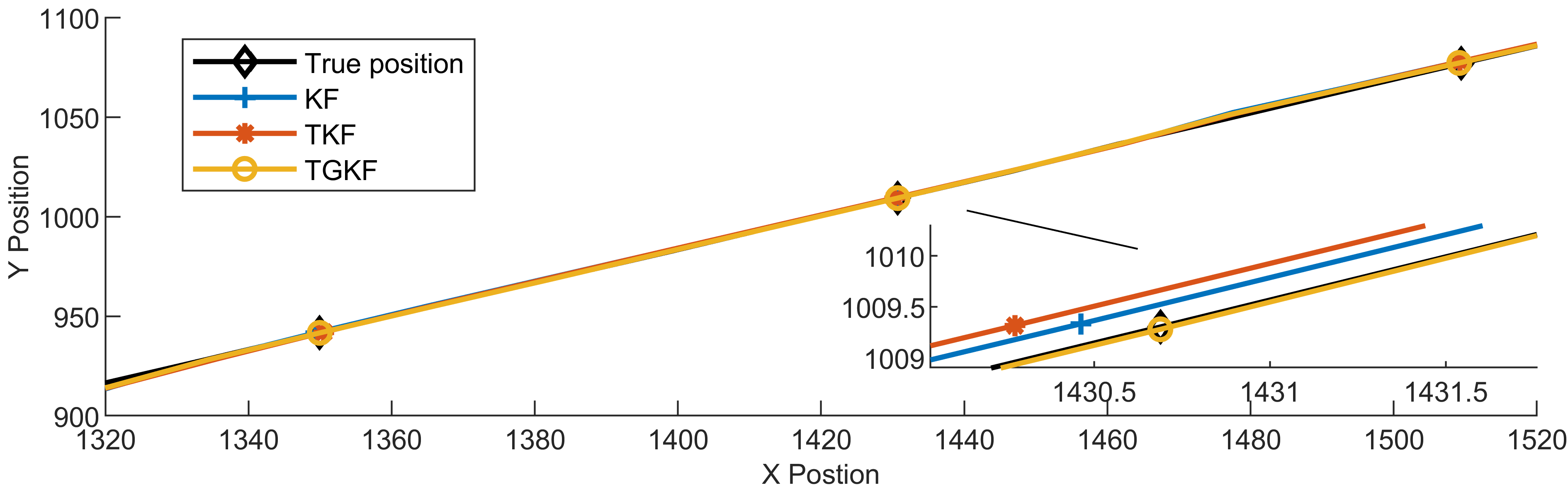}
\caption{Simulation result of trajectories.}
\label{fig_3}
\end{figure}

\begin{figure}[!h]
\centering
\includegraphics[width=3in]{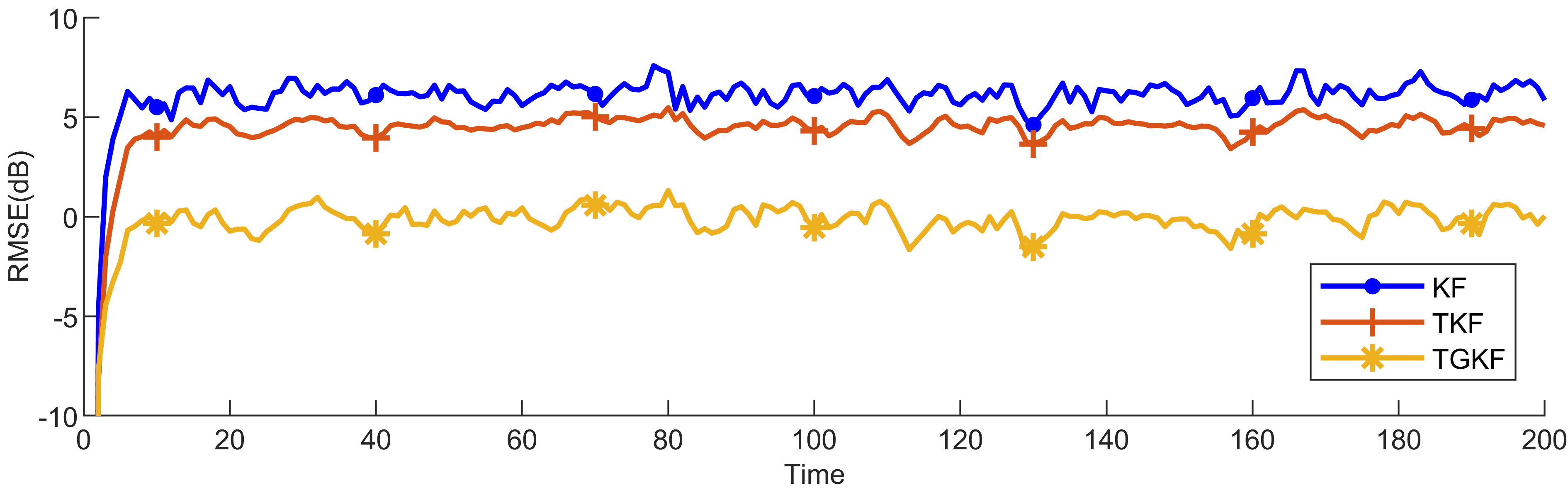}
\caption{Simulation result of RMSEs.}
\label{fig_4}
\end{figure}

It can be seen from Fig. 3 that the simulated trajectory of TGKF curves much better than the others, in meanwhile, from Fig. 4, we can see that TGKF has smaller RMSEs.

We also simulated these filters under different Gaussian percentages, under different standard deviations of measurement noises, and under alpha stable noise conditions to examine whether TGKF is robust or not, and the results are shown below:

\begin{figure}[!h]
\centering
\includegraphics[width=3in]{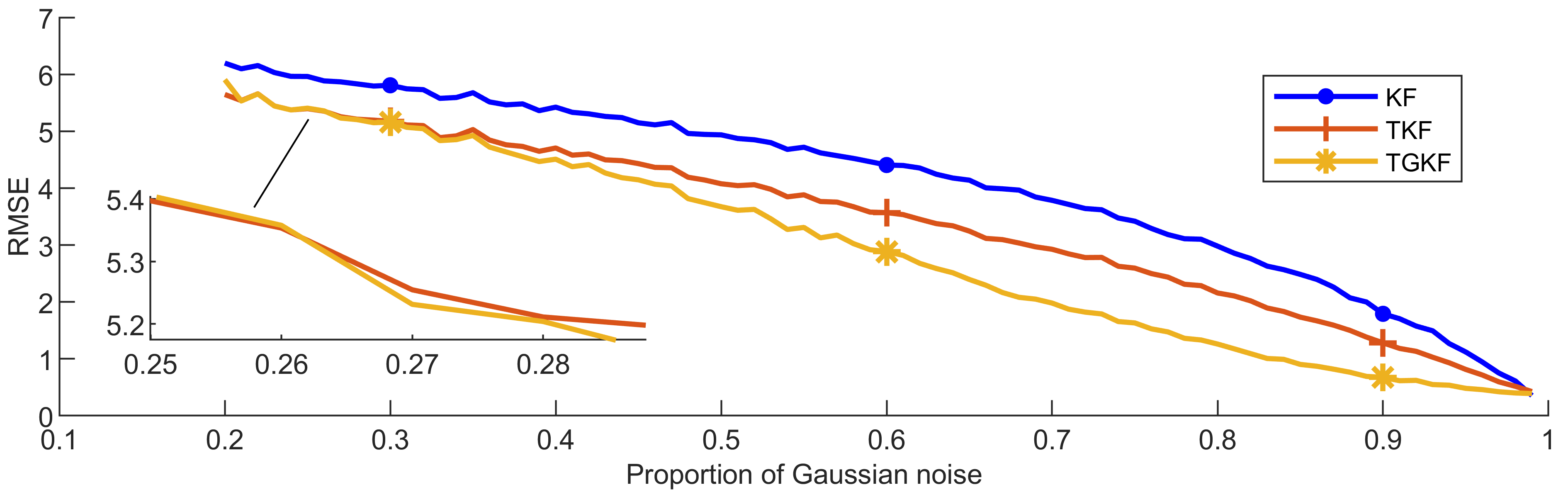}
\caption{Simulation result of Gaussian percentages.}
\label{fig_5}
\end{figure}

From Fig. 5, the filtering effect of TGKF is better than TKF and KF when the Gaussian proportion is over 30$\%$, there are only very few applications and experiments where the non-Gaussian component of the measurements can exceed 70$\%$. In such a context, TGKF is more favorable.

\begin{figure}[!h]
\centering
\includegraphics[width=3in]{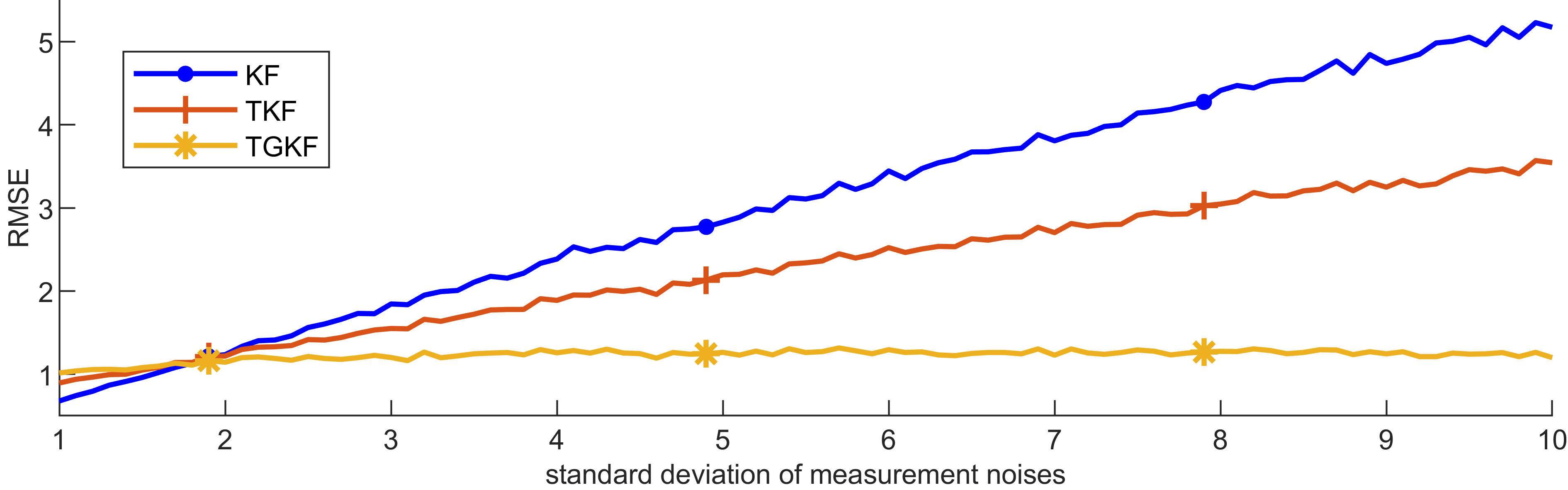}
\caption{Simulation result of standard deviations.}
\label{fig_6}
\end{figure}

From Fig. 6, TGKF has very stable RMSEs when facing larger standard deviations, and the standard deviation of impulse noise is almost always greater than 2, which indicates that TGKF is also more favorable.

\begin{figure}[!h]
\centering
\includegraphics[width=3in]{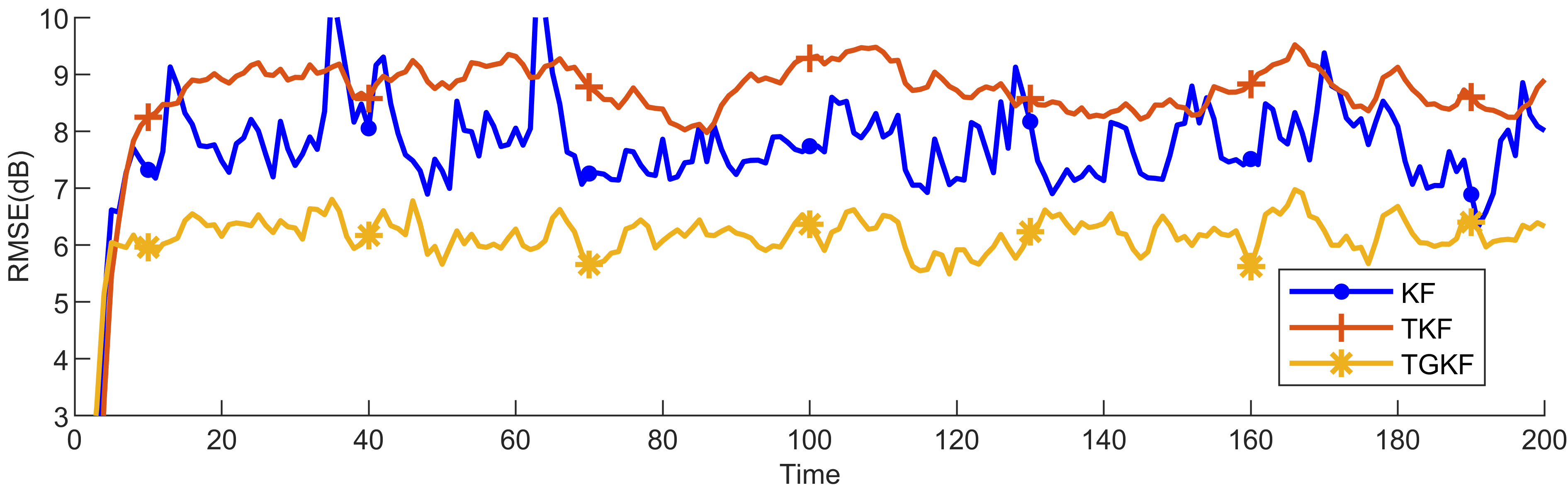}
\caption{RMSEs under alpha stable noises.}
\label{fig_7}
\end{figure}\textbf{}

From Fig. 7, TGKF can maintain good performance under alpha stable noise conditions, all the simulation results above suggests that TGKF is robust.

\section{Conclusions}

In this brief, a covariance adaptive student's t based Kalman filter was proposed, we expressed the estimation model of TKF in a simplified form and analyzed the adjustment limit of TKF when facing Gaussian noise, i.e. Equations (9)-(14), then GMM is applied to get $\textbf{\textit{R}}_\textit{k}(\textit{S})$ and $\textbf{\textit{R}}_\textit{k}(\textit{B})$ from measurement noise, combined them with the adjustment factor $\textit{TG}$, we finally get $\textbf{\textit{R}}_\textit{k}(\textit{I})$ to replace the one used in TKF. This further improved the confidence level of the measurement vector in the final estimation result when noises are Gaussian while holding the performance under non-Gaussian conditions.

TGKF significantly improved the filtering effect under Gaussian noise conditions, when it comes to practical situations, the Gaussian part often takes a high proportion in measurement noise, and thus the overall filtering effect of TGKF is better than the existing filters in application and experiment uses.

\newpage

\end{document}